\documentclass[12pt]{iopart}
\usepackage{color}
\usepackage{graphicx}
\usepackage{enumerate}
\usepackage{amssymb}
\usepackage{amsfonts}

\newcommand{\dd}{\mathrm{d}}

\begin{document}
\title[Work fluctuations in a nematic liquid crystal]{Work fluctuations in a nematic liquid crystal}
\author{S. Joubaud, G. Huillard, A. Petrossyan, S. Ciliberto}
\address{Laboratoire de Physique de l'ENS Lyon, CNRS UMR 5672,
     46, All\'ee d'Italie, 69364 Lyon CEDEX 07, France}
\ead{sylvain.joubaud@ens-lyon.fr, guillaume.huillard@ens-lyon.fr, artyom.petrossyan@ens-lyon.fr and sergio.ciliberto@ens-lyon.fr}
\begin{abstract}
The orientation fluctuations of the director of a liquid crystal
are measured, by a sensitive polarization interferometer, close to
the Fr\'eedericksz transition, which is a second order transition
driven by an electric field. Using mean field theory, we define
the work injected into the system by a change of the electric
field and we calibrate it using Fluctuation-Dissipation Theorem.
We show that the work fluctuations satisfy the Transient
Fluctuation Theorem. An analytical justification of this result is
given. The open problems for the out of equilibrium case are
finally discussed.
\end{abstract}


\pagestyle{plain}

\section{Introduction}
The physics of small systems used in nanotechnology and biology has recently received
an increased interest. In such systems, fluctuations of the work injected become of the order of the mean value,
leading to unexpected and undesired effects. For example, the instantaneous energy transfer can flow from a
cold source to a hot one. The probabilities of getting positive and negative injected work are quantitatively
related in non equilibrium system using Fluctuation Theorem (FT)~\cite{GallavottiCohen95}-\cite{Harris2006}.
This theorem has been both theoretically and experimentally studied in Brownian systems described
by a Langevin equation~\cite{Blickle2006}-\cite{Taniguchi}. The {\em Transient Fluctuation
Theorem}(TFT)
of the injected work $W_\tau$ considers the work done on the system in the transient state, {\em i.e.}
considering a time interval of duration $\tau$ which starts immediately after the external force has
been applied to the system. For these systems, FT holds for all integration time and all fluctuation magnitudes :
\begin{equation}
S(a) = \ln\left[\frac{p(W_\tau = +a)}{p(W_\tau=-a)}\right] = \frac{a}{k_BT}\qquad \forall \; a \qquad \forall \; \tau
\end{equation}
where $k_B$ is the Boltzmann constant,
$T$ the temperature of the heat bath and $p(W_\tau)$ is the probability density function (PDF)
of the injected work $W_\tau$. $S$ is called symmetry function. In a recent paper~\cite{datta2008},
some surprising results have been obtained for the TFT
in a spatially extended system where the fluctuations of the work injected by an electric
field into a nematic liquid crystal (LC) have been studied. More precisely,
 the authors find an agreement between experimental results and TFT only for
 particular values of the observation time $\tau$. We report in this article experimental results
 on the same system using a different measurement technic. We show that in our experiment,
 TFT holds for all integration time and all fluctuation magnitude. The paper is organized as follow.
 In the section~\ref{sec:exp}, we describe the experimental setup. In the section~\ref{sec:work},
 the work and the free energy of the LC system are defined. The experimental results are presented in section~\ref{sec:results}.
  Finally, we conclude in section~\ref{sec:conc}.

\section{Experimental setup\label{sec:exp}}
\begin{figure}
\centerline{\includegraphics[width=0.7\linewidth]{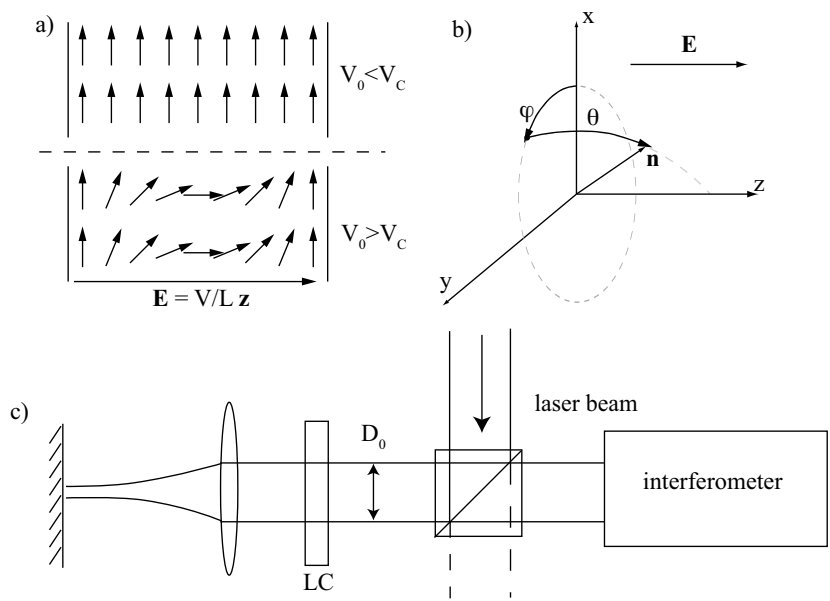}}
\caption{a) The geometry of Fr\'eedericksz transition: director
configuration for $V_0 < V_c$ and director configuration for $V_0
V_c$. b) Definition of angular displacements ($\theta$,
$\varphi$) of the director of one nematic $\vec{n}$. c)
Experimental setup. A laser beam passes through the LC cell and a
polarization interferometer measures the phase shift $\Phi$
between the ordinary an extraordinary rays~\cite{Bellon02}.}
\label{fig1:experimental_setup}
\end{figure}

In our experimental apparatus, we measure the spatially averaged alignment of the LC molecules,
whose local direction of alignment is defined by the unit vectors $\vec{n}$.
The LC is confined between two parallel glass plates at a distance $L$. The inner surfaces of the confining plates have transparent Indium-Tin-Oxyde (ITO) electrodes,
used to apply an electric field $\vec{E}$.
The plates surfaces are coated by a thin layer of polymer which is mechanically rubbed into one direction.
This treatment of the glass plates align the molecules of the LC in a unique direction parallel to the surface (planar alignment),
{\em i.e.} all the molecules near the surfaces have the same director parallel to $x$-axis and $\vec{n} =(1,0,0)$
(see figure~\ref{fig1:experimental_setup}a). This kind of coating excludes
weak anchoring effect described in the literature~\cite{cognard1982}.
The experiments are performed on nematic liquid crystal (5CB) produced by Merck in a cell of thickness $L\,=\,9$~$\mu$m.

The LC is submitted to an electric field perpendicular to the
plates by applying a voltage $V$ between the ITO electrodes. In
order to avoid electrohydrodynamic effects of the motion of the
ions invariably present in the liquid crystal, we apply an AC
voltage at a frequency of $f_V=1$~kHz ($V=\sqrt{2}V_0\cos(2\pi
f_Vt)$)~\cite{DeGennes, Oswald}. When the voltage $V_0$ exceeds a
threshold value $V_c$, the planar state becomes unstable and the
LC molecules try to align parallel to the electric field
(fig.~\ref{fig1:experimental_setup}a). This is known as the
Fr\'eedericksz transition which is second order. Above the
critical value of the field, the equilibrium structure of the LC
depends on the reduced control parameter defined as $\epsilon =
\frac{V_0^2}{V_c^2}-1$. The sample is thermostated at  room
temperature, {\em i.e.} $T = 295$~K. The stability of the
temperature is { better} than  $0.5$~K over the duration of the
experiment.

The measurement of the alignment of the LC molecules uses the
anisotropic optical properties of the LC, {\em i.e.} the cell is a
birefringent plate whose local optical axis is parallel to the
director $\vec{n}$. This optical anisotropy can be precisely
estimated by measuring the phase shift $\Phi$ between two linearly
polarized beams which cross the cell, one polarized along $x$-axis
(ordinary ray) and the other along the $y$-axis (extraordinary
ray). The experimental set-up employed is schematically shown in
fig. \ref{fig1:experimental_setup}c. The beam  produced by a
stabilized He-Ne laser ($\lambda = 632.8$~nm) passes through the
LC cell and it  is finally back-reflected into the LC cell by a
mirror. The lens between the mirror and the LC cell is used to
compensate the divergency of the beam inside the interferometer.
Inside the cell, the laser beam is parallel and has a  diameter of
about $2$~mm. The beam is at normal incidence  to the cell and
linearly polarized at $45^{\circ}$ from the $x$-axis, {\em i.e.},
can be decomposed in an extraordinary beam and in an ordinary one.
The optical path difference, between the ordinary and
extraordinary beams, is measured by a very sensitive polarization
interferometer~\cite{Bellon02}.

The phase shift $\Phi$ takes into account the angular displacement
in $xz$-plane, $\theta$, and in $xy$-plane, $\varphi$ (fig.~\ref{fig1:experimental_setup}b).
For $V_0$ larger than the threshold value $V_c$, $\Phi$ depend only on $\theta$ and can be written as :
\begin{equation}
\Phi = \frac{1}{\mathcal A}\int\hspace*{-0.2cm}\int_{\mathcal A}\dd x\dd y\left[\frac{2 \pi}{\lambda} \int_{0}^{L}
\left(\frac{n_o n_e}{\sqrt{n_0^2 \cos(\theta)^2 + n_e^2
\sin(\theta)^2}}-n_0\right) \mathrm{d} z\right]
\label{eq:phase_shift_def}
\end{equation}
with ($n_o$, $n_e$) the two anistotropic refractive
indices~\cite{DeGennes, Oswald} and $\mathcal A=\pi D_0^2/4$ is
the area of the measuring region of diameter $D_0$ in the ($x$,
$y$) plane. The global variable of our interest is $\zeta$ defined
as the spatially averaged alignment of the LC molecules, and more
precisely :
\begin{equation}
\zeta = \frac{2}{L}\frac{1}{\mathcal{A}}\int\hspace*{-0.2cm}\int_{\mathcal{A}} \dd x\dd y \int_0^{L}(1-n_x^2)\dd z
\label{eq:def_zeta}
\end{equation}
Close to the critical value $V_c$, using boundary conditions, the
space dependance of $\theta$ along $z$ has the following form :
$\theta = \theta_0(x,y,t)\sin(\pi z/L)$, thus
$n_x=\cos\theta$~\cite{DeGennes, Oswald, SanMiguel1985}. { If
$\theta_0$ remains small, $\zeta$ takes a simple form in terms of
$\theta_0$ :
\begin{equation}
\zeta = \frac{1}{\mathcal{A}}\int\hspace*{-0.2cm}\int_{\mathcal{A}} \dd x\dd y \theta_0^2
\end{equation}
After some algebra, we can show that the phase shift is a linear function of $\zeta$ :
\begin{eqnarray}
\Phi = \Phi_0 \left(1-\frac{n_e(n_e+n_o)}{4 n_o^2}\zeta\right)\\
\Phi_0 \equiv \frac{2 \pi}{\lambda} (n_e-n_o)L
\label{eq:order_parameter}
\end{eqnarray}}
The phase $\Phi$, measured by the interferometer,  is acquired
with a resolution of $24$ bits at a sampling rate of $16384$~Hz
and then filtered at $500$~Hz in order to suppress the AC voltage
at $f_V$. The need of such a high acquisition frequency will be
explained in the next section. The instrumental noise of the
apparatus~\cite{Bellon02} is  three orders of magnitude smaller
than the amplitude $\delta \Phi$ of the fluctuations of $\Phi$
induced by the thermal fluctuations of $\zeta$. In a recent paper,
we have shown that $\zeta$ is characterized by a mean value
$\langle \zeta \rangle \propto \epsilon$ and fluctuations which
have a lorentzian spectrum~\cite{Galatola, joubaud2008}.

\begin{figure}
\centerline{\includegraphics[width=0.7\linewidth]{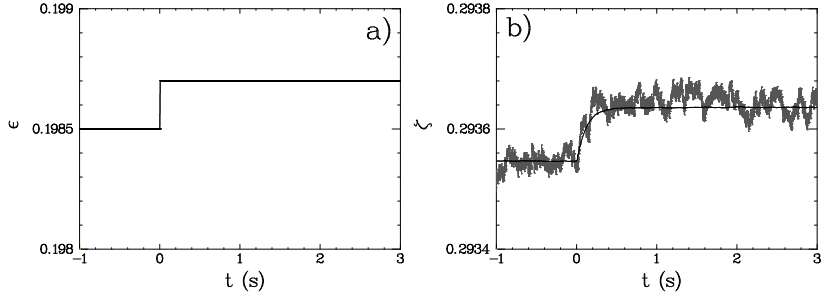}}
\caption{ Typical measurement cycle. a) Dependence of $\epsilon$
on time. The perturbation $\delta \epsilon$ is switched on at
$t=0$. b) One realization of the response of $\zeta$ when the
amplitude of the control parameter is changed from $\epsilon_0 =
0.1985$ and $\delta \epsilon = 2\,10^{-4}$ as shown in a).  The
thin line represents the average response $\langle \zeta \rangle$
done over 2500 experiments.} \label{fig:time_series}
\end{figure}

We are interested here in the fluctuations of the work injected
into the LC when it is driven away from an equilibrium state to
another using a small change of the voltage ({\em i.e.} the
control parameter), specifically:
\begin{eqnarray}
\epsilon = \cases{\epsilon_0 \quad  t\; <\; 0\\\epsilon_0+\delta
\epsilon \quad  t\; \geq 0} \label{eq:change}
\end{eqnarray}
We take care that the system is in equilibrium at $t = 0$ by
applying the value $\epsilon_0$ for a time much larger than the
relaxation time ($\tau_r$) of the system. A typical measurement
cycle is depicted in Fig.~\ref{fig:time_series} where we plot
$\epsilon(t)$ (Fig.~\ref{fig:time_series}a) and $\zeta(t)$
(Fig.~\ref{fig:time_series}b) as functions of time. We repeat the
same cycle 2500 times to compute, over this ensemble of
experiments, the probability density function of the work $W_\tau$
injected during a time $\tau$. We choose the amplitude of $\delta
\epsilon$ sufficiently small such that the mean response of the
system to this excitation is comparable to the thermal noise
amplitude, as can be seen in figure~\ref{fig:time_series}b) for
$\epsilon_0 = 0.1985$ and $\delta \epsilon = 2\,10^{-4}$.

\section{Landau equation and definition of the work\label{sec:work}}
We define now the work injected into this system using the kind of
forcing plotted in fig.~\ref{fig:time_series}a. The free energy is the sum of the electric energy and the potential energy :
\begin{equation}
F=\int\hspace*{-0.2cm}\int\hspace*{-0.2cm}\int_{\mathcal V} \dd x \dd y \dd z (f_{elastic}+f_{electric})
\end{equation}
Elastic free energy is a function of the director $\vec{n}$
and its spatial derivative~\cite{DeGennes, Oswald} and electric free energy
is $f_{elect}=-1/2\epsilon_0 \epsilon_a (\vec{E}\cdot\vec{n})^2$ where $\epsilon_a$
is the dielectric anisotropy of the LC ($\epsilon_a$ is positive for 5CB).
As the area of the measuring beam is much larger than the correlation length,
we can neglect spatial derivative over $x$ and $y$. Using the sinusoidal approximation for
$\theta$,~{\em i.e.} $\theta(x,y,z,t)=\theta_0(x,y,t)\sin(\pi z/L)$ \ \cite{DeGennes}, free energy has the following form :
\begin{eqnarray}
F &=& \int\hspace*{-0.2cm}\int_{\mathcal A}\dd x \dd y \left(\frac{\pi^2K_1}{4L}(\theta_0^2+\frac{\kappa}{4}\theta_0^4)-\frac{\epsilon_0\epsilon_a}{4L}V^2(\theta_0^2-\frac{1}{4}\theta_0^4)\right)\\
F &=&\int\hspace*{-0.2cm}\int_{\mathcal A}\dd x \dd y \left(\frac{\pi^2K_1}{4L}\left[-\epsilon\theta_0^2+\frac{1}{4}(\kappa+\epsilon+1)\theta_0^4\right]\right)
\end{eqnarray}
where $\kappa$ is equal to $(K_3-K_1)/K_1$,  $\epsilon =
V_0^2/V_c^2-1$ and $K_1$, $K_2$ and $K_3$ are the three elastic
constants of the LC. To simplify the analysis the dependance of
$\theta_0$ in $x$ and $y$ can be neglected and  the free energy
takes the simple form:
\begin{eqnarray}
F =
B\left[-\epsilon\theta_0^2+\frac{1}{4}(\kappa+\epsilon+1)\theta_0^4\right]
\label{eq:free}
\end{eqnarray}
where $B={\mathcal A} \pi^2K_1/4L$. From eq.\ref{eq:free} we can
obtain the equation of momentum~\cite{SanMiguel1985}:
\begin{eqnarray}
\frac{\gamma{\mathcal A}L}{2} \frac{{\rm d} \theta_0}{{\rm d} t}=
B\left[ \ 2\ \epsilon \ \theta_0 - (\kappa+\epsilon+1)\ \theta_0^3
\right] +\eta \label{momentum_equation}
\end{eqnarray}
where $\gamma$ is a viscosity coefficient and $\eta$ a thermal
noise delta-correlated in time, such that :
\begin{equation}
\langle \eta(t) \rangle = 0 \quad {\rm and} \quad \langle
\eta(t_1)\eta(t_2)\rangle = k_BT\gamma{\mathcal A}L\delta(t_2-t_1)
\end{equation}
where $\langle \cdot \rangle$ stands for ensemble average. First,
we have to calibrate the system, {\em i.e.} determine the value of
the constant $B$ that is a function of the area ${\mathcal A}$
which is very difficult to precisely determine on the experiment.
For this calibration, we use Fluctuation-Dissipation Theorem
(FDT). We apply a small change of the voltage $V_0$ and, so of the
control parameter, {\em i.e.} $\epsilon = \epsilon_0+\delta
\epsilon$. We separate $\theta_0$ into the average part at
$\epsilon_0$ and a deviation due to $\delta \epsilon$ :
$\theta_0(t) =\psi_0+\Delta(t)$, where $\psi_0=2 \epsilon_0 /
(\kappa+\epsilon_0+1) $ is the stationary solution of
eq.\ref{momentum_equation} at $\epsilon=\epsilon_0$. If the
response is linear, the average value of $\Delta$ is of the order
of $\delta \epsilon$. In this limit, the equation of momentum can
be rewritten :
\begin{equation}
\hspace*{-2cm}
\frac{\gamma{\mathcal A}L}{2} \frac{{\rm d} \Delta}{{\rm d} t}=  B
\left[ 2 \ \epsilon_0 - (\kappa+\epsilon_0+1)3 \psi_0^2 \right] \
\Delta + \\ 2 \ B \ \delta \epsilon \ \psi_0 \left(
1-\frac{\psi_0^2}{2}\right) + \eta + O(\Delta^2) \label{eq:Delta}
\end{equation}
The external torque, and so the conjugate variable to $\theta_0$,
is equal to \begin{equation}\Gamma_{ext} = 2B \ \delta \epsilon \ \psi_0 \
\left(1-\frac{\psi_0 ^2}{2}\right).\end{equation} We define the integrated
linear response function using an heaviside for $\delta \epsilon$
:
\begin{equation}
R(\tau) = \frac{\langle \Delta(\tau) \rangle}{\Gamma_{ext}} =
\frac{1}{2B\ \psi_0 \ \left(1-\frac{\psi_0^2}{2}\right)} R_{\delta
\epsilon}
\end{equation}
where $R_{\delta \epsilon}$ is the integrated response function of
$\theta_0$ to $\delta \epsilon$. $R$ is related by FDT to the
autocorrelation function,$C_{\theta_0}(\tau) = \langle
\delta\theta_0(t+\tau)\ \delta \theta_0(t)\rangle$, of the
fluctuations $\delta \theta_0$ of $\theta_0$, that is:
\begin{equation}
R(\tau) = \frac{1}{k_BT}\left(C_{\theta_0}(0) - C_{\theta_0}(\tau)\right)
\end{equation}
The global variable measured by the interferometer and  defined in
eq.~\ref{eq:def_zeta}  is $\zeta = 1/{\mathcal A}
\int\hspace*{-0.2cm}\int \theta_0^2\dd x \dd y \simeq \psi_0^2 + 2
\psi_0 \ \delta \theta_0$. Thus the mean value  of $\zeta$ is
$\zeta_0=\psi_0^2$ and the  fluctuations $\delta \zeta$ of $\zeta$
can be related to the fluctuations of $\theta_0$ : $\delta \zeta =
2 \psi_0 \delta \theta_0$. We measure the autocorrelation function
of $\delta \zeta$, $C_{\zeta} = 4\psi_0^2 C_{\theta_0}$. The
response function of $\zeta$ to $\delta \epsilon$ is also related
to $R_{\delta \epsilon}$ : $R_{\zeta,\delta \epsilon} = 2\psi_0
R_{\delta \epsilon}$. Using fluctuation-dissipation theorem, we
can measure the constant $B$ :
\begin{equation}
\frac{1}{B'}R_{\zeta,\delta\epsilon} =
\frac{1}{k_BT}\left(C_{\zeta}(0)-C_{\zeta}(\tau)\right)
\label{FDT}
\end{equation}
where $B'=B \ \left(1-\psi_0^2/ {2}\right)$.

We  define now the work, $W_\tau$, injected into the system when
the control parameter $\epsilon$ is suddenly changed from
$\epsilon_0$ to $\epsilon = \epsilon_0+\delta \epsilon$ at $t=0$
(eq.\ref{eq:change}). Using equation~\ref{momentum_equation},
$W_\tau$ can be written as :
\begin{equation}
W_\tau = B \int_0^\tau \delta \epsilon(t') \frac{{\rm d}\theta_0}{{\rm d} t}\left(2 \theta_0 -\theta_0^3\right) \dd t'
\end{equation}
we consider that the external torque is ${B \delta \epsilon(t')
\left(2 \theta_0 -\theta_0^3\right)}$. This expression can be
integrated into :
\begin{equation}
W_\tau = B \delta \epsilon\left(\theta_0^2(\tau) - \theta_0^2(0) -
\frac{\theta_0^4(\tau)-\theta_0^4(0)}{4}\right),
\end{equation}
{Using the assumption of the limit of small $\theta_0$ and
neglecting its dependance in the $x$ and $y$ directions, the work
has a simple form in terms of $\zeta=\theta_0^2$ :}
\begin{equation}
W_\tau = B \delta \epsilon\left(\zeta(\tau) - \zeta(0) -
\frac{\zeta^2(\tau)-\zeta^2(0)}{4}\right) \label{eq:work}
\end{equation}
Notice that if, $\epsilon_0$ is not too large, the quadratic term
in this expression can be neglected, but it is not the case in
general. The work $W_\tau$ is of course a fluctuating quantity
because of the fluctuation of $\zeta$.

\section{Fluctuations of the work injected into the system\label{sec:results}}
\subsection{Experimental results}
\begin{figure}
\centerline{\includegraphics[width=0.85\linewidth]{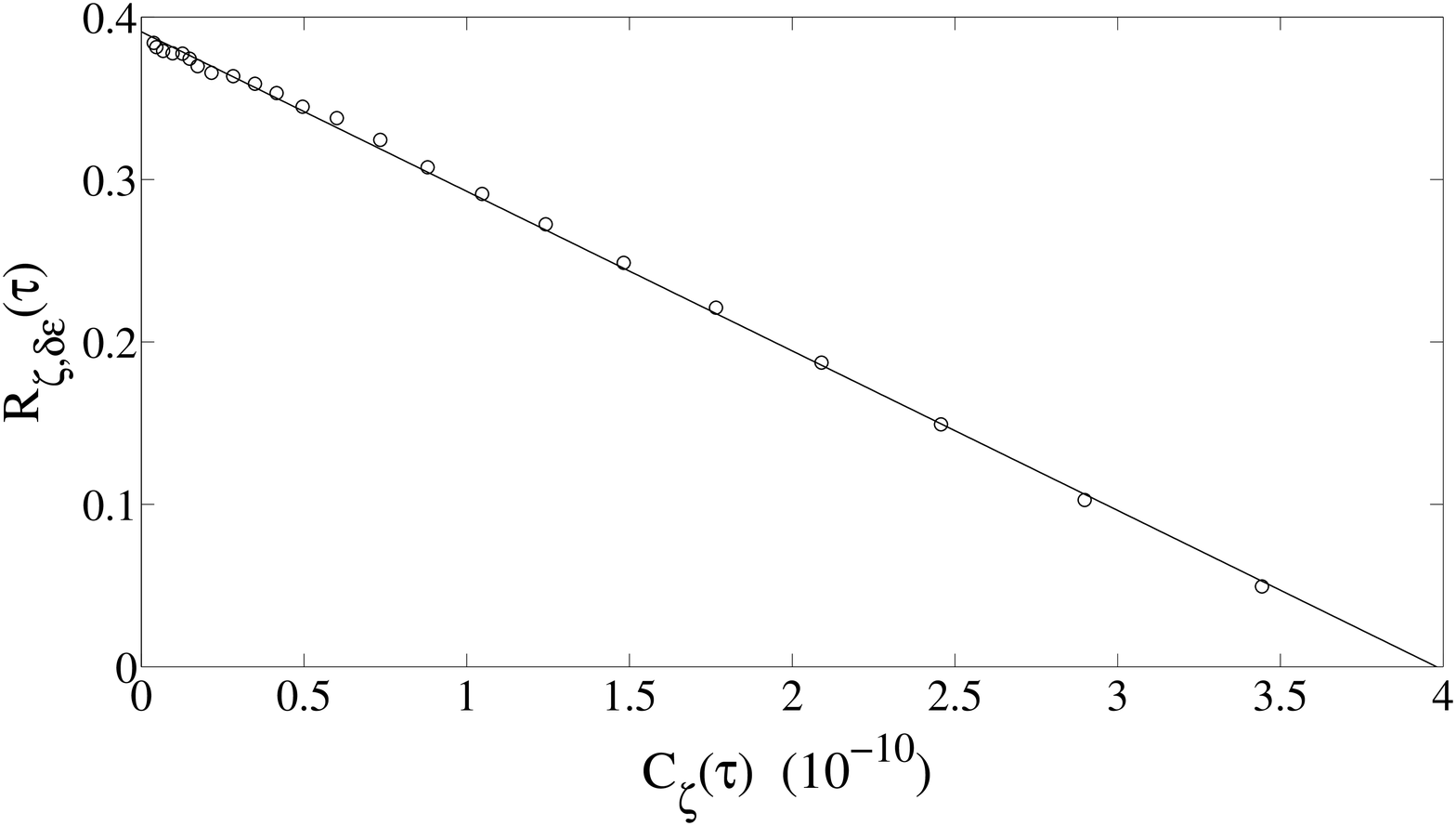}}
\caption{{\bf Fluctuation-dissipation theorem.} $R_{\zeta,
\delta \epsilon}(\tau)$ is represented as a function of
$C_{\delta \zeta}(\tau)$. Continuous line represents linear fits.
The measurement has been done at $\epsilon=0.1985$ with $\delta
\epsilon= 2\,10^{-4}$ } \label{fig:fdt}
\end{figure}

The first step consists in  calibrating  the system. Indeed as we
have already mentioned  the parameter $B'$ appearing in
eq.\ref{FDT} depends on $\mathcal A$, i.e. the effective area of
the laser beam inside the cell that is very difficult to estimate
with a good accuracy. In order to calibrate the system we use FDT.
In this case we use a sufficiently small perturbation $\delta
\epsilon$ to insure that the response is linear.  Thus we obtain
the integrated linear response of the system : $ R_{\zeta,\delta
\epsilon}(\tau)$. Then  we measure the correlation $C_\zeta$. The
function $R_{\zeta, \delta \epsilon}(\tau)/B'$ is a linear
function of $C$, as can be seen in figure~\ref{fig:fdt} and the
slope seems to be independent of $\langle \zeta \rangle$. Using
FDT, the slope is equal to $B'/k_BT$; by this method, we can
calibrate the constant $B'$. We measure $B'=4.03\,10^{-2} \pm 2
\,10^{-3}$.   Using in the equation for
$B'=(1-\psi_0^2/2)(\pi^2K_1{\mathcal A})/(4L)$, the measured
values of $B'$ and $\psi_0$, $K_1 = 6.4\, 10^{-12}$~N (value for the
LC 5CB) and $L=9$~$\mu$m we estimate  $\mathcal A=2.4
\,10^{-6}$~m$^2$. This value  is consistent with a $D_0= 1.75$~mm
which is very close to the estimated  diameter of the laser beam
inside the cell which is about $2$~mm. This justify our approach.

\begin{figure}
\centerline{
\parbox{\linewidth}{\centerline{\includegraphics[width=1.15\linewidth]{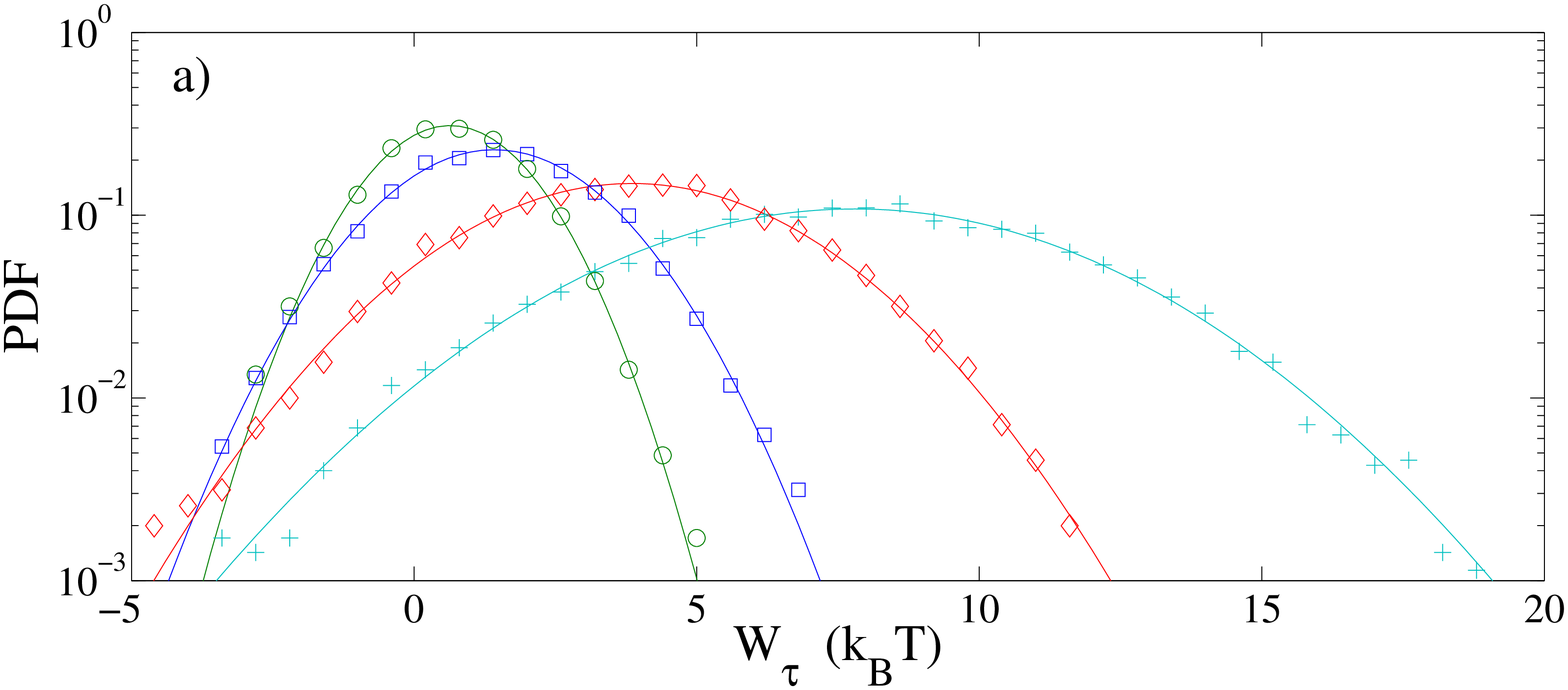}}}}
\centerline{
\parbox{\linewidth}{\centerline{\includegraphics[width=1.15\linewidth]{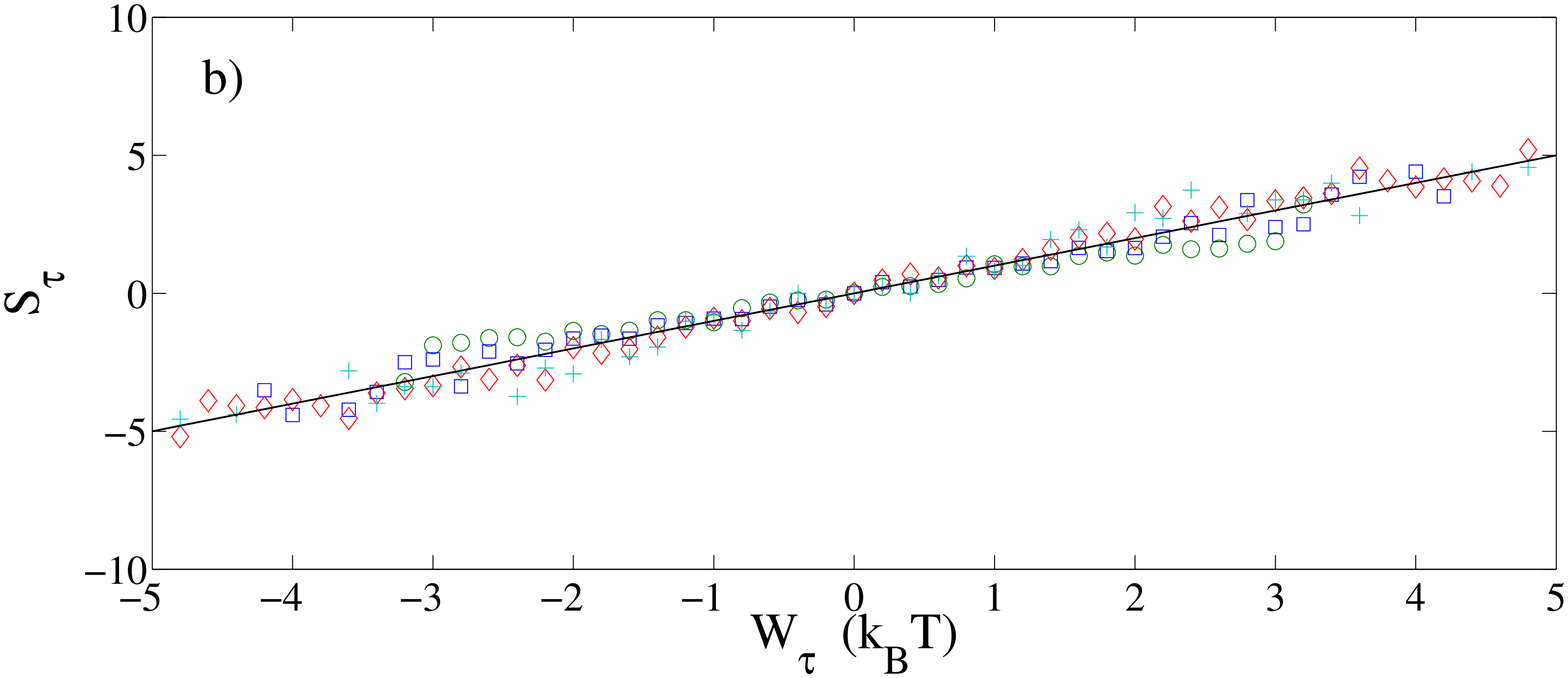}}}}
\caption{{\bf Work fluctuations} a) PDFs of the work $W_\tau$
injected into the system at $\epsilon_0= 0.1985$ by a perturbation
$\delta \epsilon= 2\, 10^{-4} $, for various $\tau$ : $10$~ms ($\circ$),
$20$~ms ($\Box$), $50$~ms ($\diamond$) and $100$~ms ($+$).
Continuous lines are Gaussian fits. b) Symmetry functions
$S(W_\tau)$ computed using the PDF of a) for the same values of
$\tau$. The straight continuous line is a line with slope $1$ for
all values of $\tau$.} \label{fig:PDF}
\end{figure}

The system is at equilibrium at $t=0$~s and we are now interested
in the work done by the change of $\delta \epsilon$ during a time
$\tau$, $W_\tau$. We need a high acquisition frequency to reduce
experimental errors in the determination of the first point of
each cycle $t=0$. The probability density function of $W_\tau$ are
plotted in figure~\ref{fig:PDF} for different values of $\tau$. We
find that the PDFs are Gaussian for all values of $\tau$ and the
average value of $W_\tau$ is equal to a few $k_B T$. We also
notice that the probability of having negative values decreases
when $\tau$ is increased. The symmetry function is plotted in
fig.~\ref{fig:PDF}b. It is a linear function of $W_\tau$ for any
$\tau$, that is $S(W_\tau) = \Sigma(\tau) W_\tau$. Within
experimental errors, we measure the slope $\Sigma(\tau) = 1$.
Thus, for our experimental system, the TFT is verified for any
time $\tau$.

\begin{figure}
\centerline{
\parbox{\linewidth}{\centerline{\includegraphics[width=0.75\linewidth]{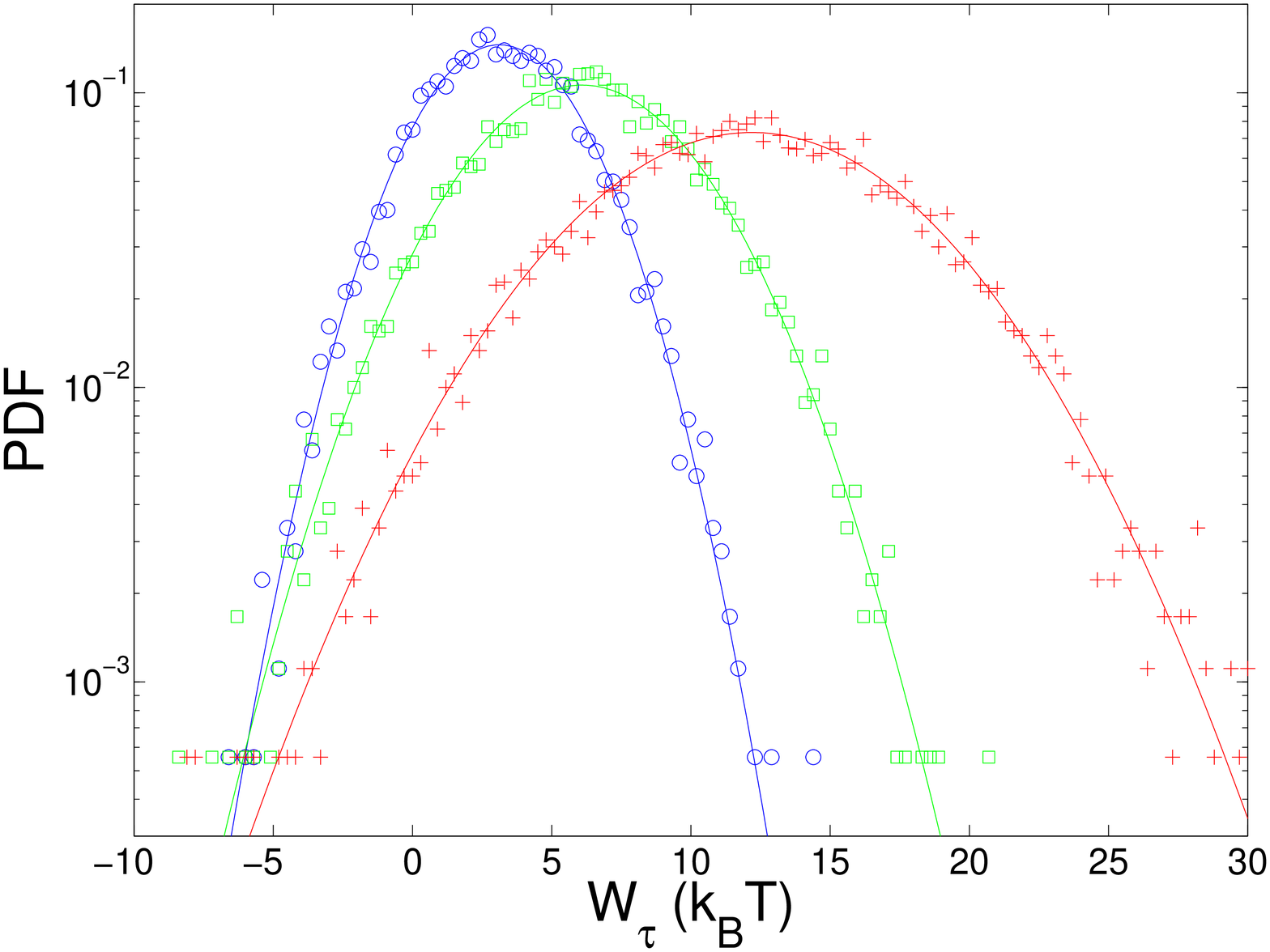}}}}
\centerline{
\parbox{\linewidth}{\centerline{\includegraphics[width=0.75\linewidth]{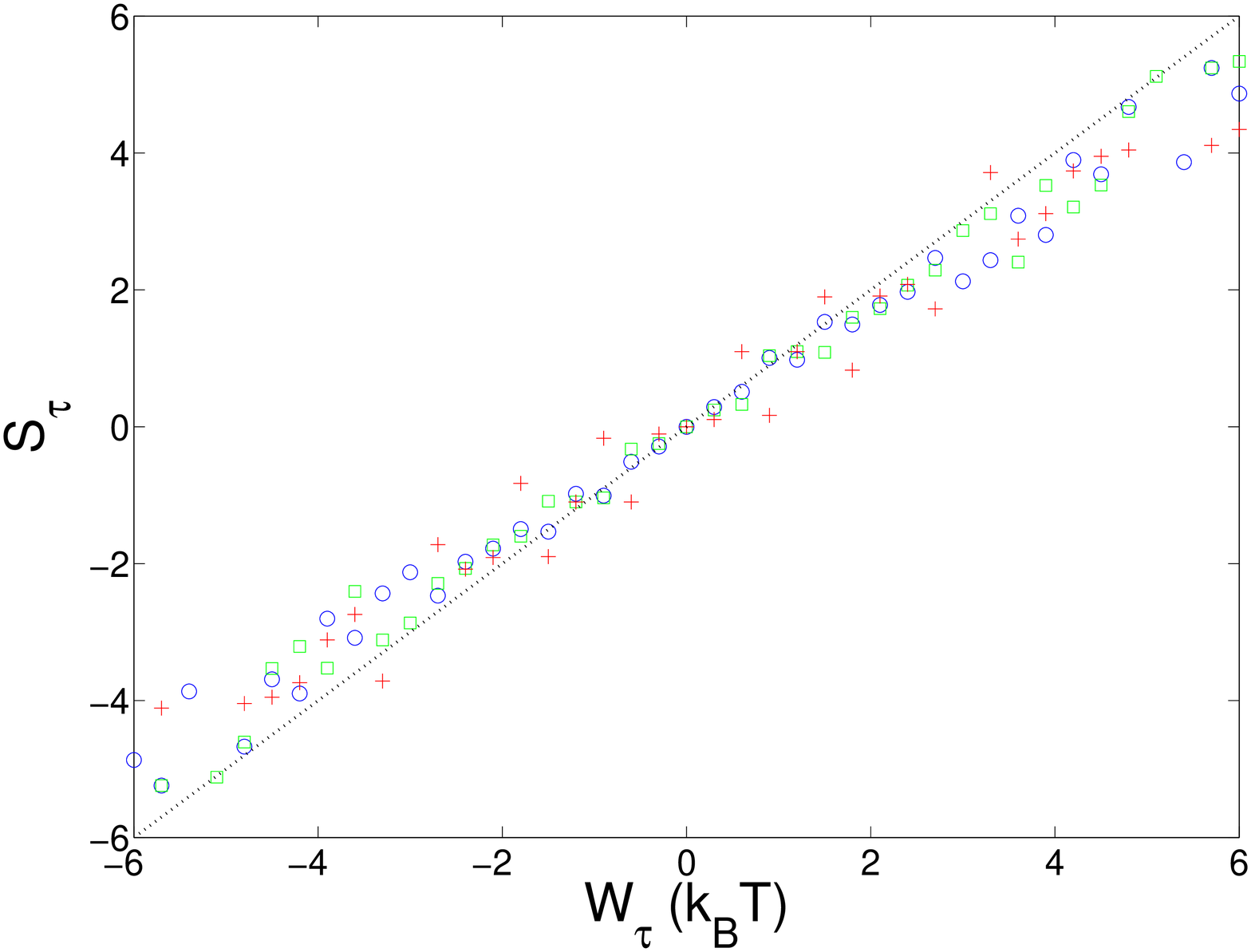}}}}
\caption{{\bf Work fluctuations} a) PDFs of the work $W_\tau$
injected into the system at $\epsilon_0= 1.039$ by a perturbation
$\delta \epsilon= 4\, 10^{-4} $, for various $\tau$ : $10$~ms ($\circ$),
$20$~ms ($\Box$), $50$~ms ($+$).
Continuous lines are Gaussian fits. b) Symmetry functions
$S(W_\tau)$ computed using the PDF of a) for the same values of
$\tau$. The dashed line is a line with slope $1$ for
all values of $\tau$.} \label{fig:PDF2}
\end{figure}

{ To test the importance of the quartic terms in the expression of
the work, we have done the same experiments using a higher value
of the control parameter $\epsilon_0=1.039$. The perturbation
remains small $\delta \epsilon = 4\, 10^{-4}$ (see
figure~\ref{fig:PDF2}). The quadratic terms in the expression of
the work cannot be neglected in this situation. We find that the
PDFs of the work are also Gaussian for all the integration times.
This is normal because the perturbation is such small that we are
in the realm of the applicability of the FDT. The symmetry
function is plotted in fig.~\ref{fig:PDF2}b. It is a linear
function of $W_\tau$ for any $\tau$. Within experimental errors,
we measure the slope equal to $1$. Thus, for our experimental
system, the TFT is verified for any time $\tau$.}

\subsection{Comparison with theory}
These results can be  justified for small $\epsilon_0$ and $\delta \epsilon$. Decomposing  $W_\tau$ into the
sum of a mean part $\langle W_\tau \rangle$ and a fluctuating one
$\delta W_\tau$, {\em i.e.} $W_\tau = \langle W_\tau \rangle +
\delta W_\tau$. The distributions of the injected work are
Gaussian for all values of $\tau$, so the symmetry function takes
a simple form :
\begin{equation}
S(W_\tau) = \frac{2\langle W_\tau \rangle}{\sigma_W^2}W_\tau
\end{equation}
where $\sigma_W^2$ is the variance of the fluctuation of the injected work. The mean value of the work is for small $\epsilon_0$ and small $\delta \epsilon$ :
\begin{eqnarray}
\langle W_\tau \rangle &=& B\delta \epsilon (\langle \zeta(\tau) \rangle - \langle \zeta(0) \rangle)\\
&=& R_{\zeta,\delta \epsilon}(\tau) B^2\delta \epsilon^2
\end{eqnarray}
The variance of the fluctuations is given by :
\begin{eqnarray}
\sigma_W^2 &=& B^2 \delta \epsilon^2 (\langle \delta \zeta(\tau)^2
\rangle + \langle \delta \zeta(0)^2\rangle - 2\langle \delta \zeta(\tau)\delta \zeta(0))\\
&=& 2 B^2 \delta \epsilon^2 (C(0) - C(\tau))
\end{eqnarray}
Using FDT, we obtain that $2k_BT\langle W_\tau \rangle =
\sigma_W^2$, thus FT is satisfied for all integration times and
all fluctuation magnitudes as can be seen in fig.~\ref{fig:PDF}b).

\section{Conclusion\label{sec:conc}}
In conclusion, we have defined in a LC cell close to
Fr\'eedericksz transition the work injected into the system by the
change of $\delta \epsilon$ during a time $\tau$ { using the
assumption of small angle $\theta_0$ and neglecting its dependance
in $x$ and $y$ directions}. We have shown that the TFT holds for
the work injected in a spatially extended system for all
integration times and all fluctuation magnitudes. We have tested
this result for another thickness of the LC cell and other values
of $\epsilon$. It is interesting to notice that this experimental
test is the first for a spatially extended system in the presence
of a non linear potential. This is in contrast to the result of
ref~\cite{datta2008}. The reason of this discrepancy is that they
neglect in the definition of the work (eq.\ref{eq:work}) the
square term in $\zeta^2$.  This cannot be done when a large
$\epsilon_0$ is used as in the case of ref.\cite{datta2008}.

Our results are  a preliminary study for future investigation on
unsolved problems such as the applicability of Transient
Fluctuation Theorem in an out-of-equilibrium system such as an
aging system or a rapid quench.

\end{document}